\documentstyle[epsf,floats,twocolumn,eqsecnum,prl,aps]{revtex}
\newcommand{\bS}{\mbox{\boldmath $S$}}
\newcommand{\sech}{{\rm sech}}
\begin{document}
\draft
\preprint{}
\title{Spectral flow of non-hermitian 
Heisenberg spin chain with complex twist}
\author{ Takahiro Fukui\cite{Email}} 
\address{Institute of Advanced Energy, Kyoto University,
Uji, Kyoto 611, Japan}
\author{Norio Kawakami}
\address{Department of Applied Physics,
Osaka University, Suita, Osaka 565, Japan} 
\date{September 29, 1997}
\maketitle
\begin{abstract}
We investigate the spectral flow of the integrable 
non-hermitian Heisenberg spin chain under 
boundary conditions with complex twist angle.
It is shown that the period of the spectral flow is
$4\pi$ up to a certain critical imaginary twist,
beyond which the period jumps successively to higher values. 
We argue that this phenomenon caused by 
non-hermitian properties of the system is closely related
to the metal-insulator transition 
caused by non-hermitian hoppings for the one-dimensional 
insulator.
\end{abstract}
\vspace{5mm}
\section{Introduction}

Boundary conditions have been playing an important role in statistical 
mechanics. 
Imagine for instance 
a one-dimensional (1D) many particle system with twisted boundary
conditions where a particle gets additional phase factor $e^{i\Phi}$ 
when it moves once along the system with a ring geometry 
and return to the previous position. 
For integrable systems, this general phase factor is 
simply incorporated into the Bethe
ansatz equations, and their characteristic properties  
have been intensively studied exactly\cite{Veg,ABB,YunBat,KBI}.
Physically, the twisted boundary 
condition is equivalent to introducing a magnetic flux threaded in the 
ring system\cite{ByeYan}. In proportion as the flux increases, 
the spectrum of the system flows in a complex way, experiencing 
many level crossings and level repulsions.
We can get detailed information for the physical properties
from the spectral flow thus obtained. For example, 
a local behavior of the spectral flow near $\Phi\sim0$ 
directly gives the transport coefficient such as  the Drude weight, 
which is a sensitive probe of the metal-insulator 
transition\cite{Koh,ShaSut}. From 
a more global structure of the spectral 
flow\cite{SutSha,YuFow,RES}, especially paying attention to its period
for the ground state, one can also determine
whether or not a bound state forms\cite{Sut,KusAok}.
Berry's phase was calculated in this process\cite{KorWu,FIO},
which may carry the information for statistics of elementary 
excitations. It was indeed shown \cite{FukKaw1,FukKaw2,LiuWan} that 
the period of the spectral flow is directly related to Haldane's 
fractional exclusion statistics \cite{Hal}.  Also, 
the twisted boundary condition has 
attracted other current interests\cite{Kit,NasTir}.

Normal twisted boundary conditions can be further generalized to 
those with complex twist angles, 
$\Phi\rightarrow\Phi+i\Psi$\cite{YunBat}.
They make the Hamiltonian non-hermitian, which is the most remarkable
difference from normal twisted boundary conditions.
The spectral flow is now defined on the complex plane and shows a rich
structure.  An interesting point to be emphasized is that this problem 
has the close relationship to recent topics studied actively 
in the context of non-hermitian systems. 
In particular, it was claimed by Hatano and Nelson 
\cite{HatNel} that introducing non-hermitian hopping terms 
into the Hamiltonian brings about the localization-delocalization 
transition even for 1D random systems.
The key point is that eigenvalues become complex 
if we increase the imaginary twist angle, which can 
provide a criterion whether the delocalization happens.
Many interesting problems on the non-hermitian Hamiltonian systems 
still remain to be explored. In order to study such problems, 
integrable non-hermitian models are quite useful.

In this paper, we investigate a 1D non-hermitian system with
twisted boundary conditions, and discuss how the spectral flow 
changes its character if  the imaginary twist is introduced.
As a typical example, we examine a version of
the asymmetric Heisenberg spin chain and the  related vertex models
studied by \cite{SYY,Nol,BukSho,GwaSpo,ADHR,NeeNij,Kim,ADW,NohKim}.
It was argued that asymmetry causes interesting  phenomena.
For example, by introducing non-hermitian XY components, 
the massive XXZ spin chain with Ising-type anisotropy 
is driven to the massless phase, which is regarded as a 
kind of the metal-insulator transition\cite{ADW}.
The model we will study in this paper is the {\it isotropic}
Heisenberg spin chain, which is massless even for zero imaginary twist. 
However, if we consider  a finite-size system,
there exist many small gaps due to finite-size effects.
It is these gaps that determine the period of the ground state in the
global spectral flow. We show that introducing the 
non-hermitian imaginary twist may collapse these gaps, 
changing the period to a larger one. It is  argued that
this phenomenon has essentially the
same origin as for the bulk metal-insulator
transition in 1D non-hermitian many-body systems:
In both cases, the transition is triggered by complex eigenvalues 
due to non-hermitian hopping terms.

This paper is organized as follows. In the next section, we 
introduce the model and the Bethe ansatz
equations, and briefly summarize the characteristics in 
the spectral flow of the ordinary symmetric (hermitian)
model. In section III, we discuss the spectral flow of the asymmetric
model with complex twist angles, 
by numerically evaluating the Bethe-ansatz spectrum for small
systems. We show that there is indeed the critical value
of complex twist, at which the period of the spectral flow suddenly
changes. In IV, we estimate this critical value for
the transition for large but finite-size systems. For this purpose, 
we propose an approximate treatment to 
obtain the finite-size corrections rather accurately.
Final section V is devoted to summary and discussions.

\section{Model and basic properties of the spectral flow}

In this section, we introduce the model and summarize some basic 
properties of the spectral flow for the ground state of the 
hermitian (symmetric) model. 
\subsection{Model and Bethe ansatz equations}
The model we consider is the well-known isotropic Heisenberg spin chain
\begin{equation}
H=\sum_{j=1}^N\left(\bS_j\cdot\bS_{j+1}-\frac{1}{4}\right),
\label{Ham}
\end{equation}
but with the twisted boundary condition,
\begin{equation}
S_{j+N}^{\pm}=e^{\pm i\Theta}S_j,\quad S_{j+N}^z=S_j^z.
\end{equation}
The spectral flow has been discussed so far for real $\Theta$.
What we are now interested in is how the ground state behaves for 
the complex twist angle,
\begin{equation}
\Theta=\Phi+i\Psi,
\label{TwiAng}
\end{equation}
which makes the Hamiltonian (\ref{Ham}) non-hermitian.
In the case of purely imaginary twist angle, the model is the same as
the asymmetric Heisenberg chain but with the replacement 
$\Psi\rightarrow N\Psi$. 
The asymmetric XXZ Heisenberg chain has been studied so far
in the massive regime, while our isotropic model is in the
massless regime for which the non-hermitian  effects disappear 
in the thermodynamic limit.  We shall in turn treat the model for
a finite-size system, and discuss how the non-hermitian 
property affects its spectral flow by examining the $1/N$  corrections.

We begin by summarizing the symmetry properties of the model.
Define the eigenvalue problem as
\begin{eqnarray}
&&H(\Theta)|\psi(\Theta)\rangle_R=~E_R(\Theta)|\psi(\Theta)\rangle_R
\nonumber\\
&&_L\langle\psi(\Theta)|H(\Theta)= ~ _L\langle\psi(\Theta)|E_L(\Theta)
\end{eqnarray}
Because of the symmetries $H^*(\Theta)=H(-\Theta^*)$,
$H^\dagger(\Theta)=H(\Theta^*)$, and
$UH(\Theta)U^\dagger=H(-\Theta)$ with $U=e^{-i\pi\sum S_j^y}$
we have 
\begin{eqnarray}
&&|\psi(-\Theta^*)\rangle_R=|\psi^*(\Theta)\rangle_R,\quad
E_R(-\Theta^*)=E_R^*(\Theta),
\nonumber\\
&&|\psi(\Theta^*)\rangle_R=U|\psi^*(\Theta)\rangle_R,\quad
E_R(\Theta^*)=E_R^*(\Theta),
\nonumber\\
&&|\psi(\Theta)\rangle_L=U|\psi(-\Theta^*)\rangle_R,\quad
E_L(\Theta)=E_R(\Theta).
\end{eqnarray}

As shown by Yung and Batchelor\cite{YunBat}, the model is solved by
the Bethe ansatz method,
\begin{equation}
Np(\lambda_j)+\sum_{k\ne j}^M\theta(\lambda_j-\lambda_k)
=2\pi I_j+\Theta, \quad j=1,\cdots,M,
\label{BAE}
\end{equation}
where the momentum $p(\lambda)$ and the 
two-body phase shift $\theta(\lambda)$
as functions of the rapidity $\lambda$ are given by
\begin{eqnarray}
&&p(\lambda)=-i\ln\left(-\frac{\lambda-i/2}{\lambda+i/2}\right),
\nonumber\\
&&\theta(\lambda)=i\ln\left(-\frac{\lambda-i}{\lambda+i}\right).
\label{MomPha}
\end{eqnarray}
Here the branch cut lies on $(-i\infty,-i/2)$ and $(i/2,i\infty)$ 
for $p(\lambda)$ and on $(-i\infty,-i)$ and $(i,i\infty)$ for
$\theta(\lambda)$.
The total energy of the system is given by the summation of single 
particle energies,
\begin{equation}
E=\sum_{j=1}^M\frac{-2}{4\lambda_j^2+1}.
\end{equation}
In the rest of the paper, we concentrate on the spectral flow of
{\it the ground state}, so that we set $M=N/2$ with $N=$ even, and
\begin{equation}
I_j: -\left(\frac{N}{4}-\frac{1}{2}\right),\cdots,
\left(\frac{N}{4}-\frac{1}{2}\right).
\end{equation}

\subsection{Spectral flow for the case $\Psi=0$}
\begin{figure}[htb] 
\epsfxsize=70mm 
\centerline{\epsfbox{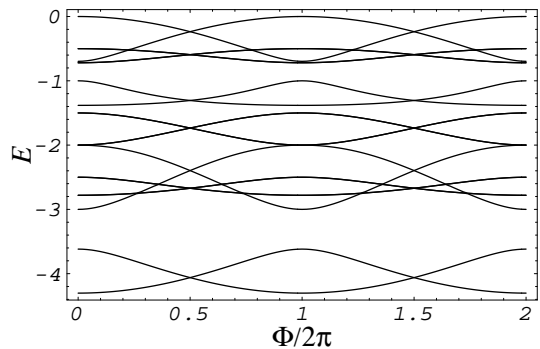}} 
\vspace{0.2cm}
\caption{Spectral flow of the $S^z=0$ sector of the 
$N=6$ with $\Psi=0$
system obtained by the exact numerical diagonalization}
\label{f:diag6_00}
\end{figure}

Before we proceed to the non-hermitian model, let 
us briefly  summarize the
spectral flow of the model with an ordinary hermitian 
condition $\Psi=0$. This case was investigated in detail 
by \cite{ABB,SutSha,YuFow,KorWu,FIO}, who claimed 
that the ground state becomes the first excited state at $\Phi=2\pi$
and returns to the ground state again at $\Phi=4\pi$, namely, the period
of the ground state is $4\pi$ as a function of $\Phi$.
This feature can be observed indeed in Fig. \ref{f:diag6_00}, which is
obtained by the exact diagonalization 
for the $S^z=0$ sector of the $N=6$ system.
The period $4\pi$ is caused by the small gaps
made by level repulsions due to  
irrelevant backward interactions, and is quite general 
for ordinary finite-size systems with such irrelevant
interactions. This property will be 
essentially changed by the introduction of non-hermitian terms,
as is discussed in the next section.

Before concluding this section, we wish to give an interpretation 
of the above spectral flow in terms of the Bethe-ansatz language
\cite{SutSha,FIO}.
At $\Phi=0$ the $N$ rapidities for the ground state 
are distributed symmetrically on the
real axis. As $\Phi$ approaches $2\pi$, the last rapidity $\lambda_M$
runs to infinity, while the others shift a little bit to the positive
direction, and are arranged in a configuration of the lowest energy state 
for  the $S^z=1/2$ sector.
This can be seen clearly if we divide the Bethe ansatz equations
(\ref{BAE}) at $\Phi=2\pi$ into two classes,
\begin{equation}
Np(\lambda_j)+\sum_{k\ne j}^{M-1}\theta(\lambda_j-\lambda_k)
+\theta(\lambda_j-\lambda_M)
=2\pi(I_j+1), 
\end{equation}
for  $j=1,\cdots,M-1$ and 
\begin{equation}
Np(\lambda_M)+\sum_{k=1}^{M-1}\theta(\lambda_M-\lambda_k)
=2\pi(I_M+1),
\label{Img}
\end{equation}
The last equation is satisfied by $\lambda_M=+\infty$, and the former
equations reduce to
\begin{equation}
Np(\lambda_j)+\sum_{k\ne j}^{M-1}\theta(\lambda_j-\lambda_k)
=2\pi I_j',
\label{BAEReal}
\end{equation}
where
\begin{equation}
I_j': -\frac{N}{4},\cdots,\frac{N}{4}.
\end{equation}
Beyond $\Phi=2\pi$, the rapidity $\lambda_M$ jumps from 
infinity to minus 
infinity and at last all rapidities return to the original configuration
at $\Phi=4\pi$. In this process, the rapidity $\lambda_M$ crosses the
branch cut of the logarithmic function in eq.(\ref{MomPha}), so that 
\begin{eqnarray}
&&p(\lambda_j^{(4\pi)})=p(\lambda_{j+1}^{(0)}),\quad j=1,\cdots,M-1,
\nonumber\\
&&p(\lambda_M^{(4\pi)})=p(\lambda_1^{(0)})+2\pi.
\end{eqnarray}
This gives an interpretation of  the period $4 \pi$, which is due to 
irrelevant interactions, in terms of the motion of the rapidities.
In the next section, by extending the above idea,
we discuss how the spectral flow changes its character if we
switch on the imaginary twist $\Psi$.

\section{Spectral flow for the non-hermitian model}
\subsection{Basic properties of the spectral flow}

\begin{figure}[htb]
\epsfxsize=70mm 
\centerline{\epsfbox{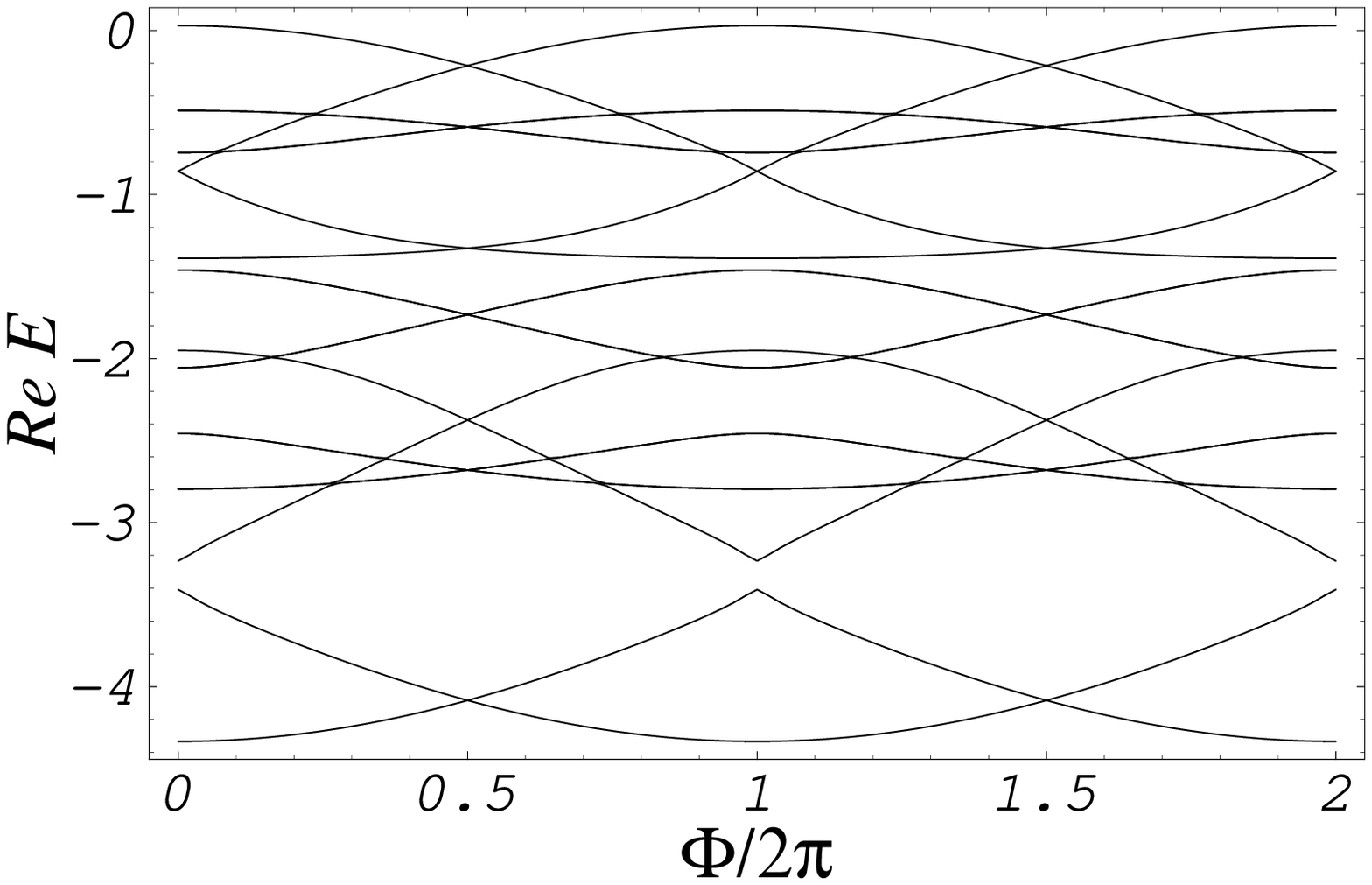}} 
\vspace{0.2cm}
\caption{The same spectral flow but with $\Psi=1.1$}
\label{f:diag6_11}
\end{figure}

\begin{figure}[htb]
\epsfxsize=70mm 
\centerline{\epsfbox{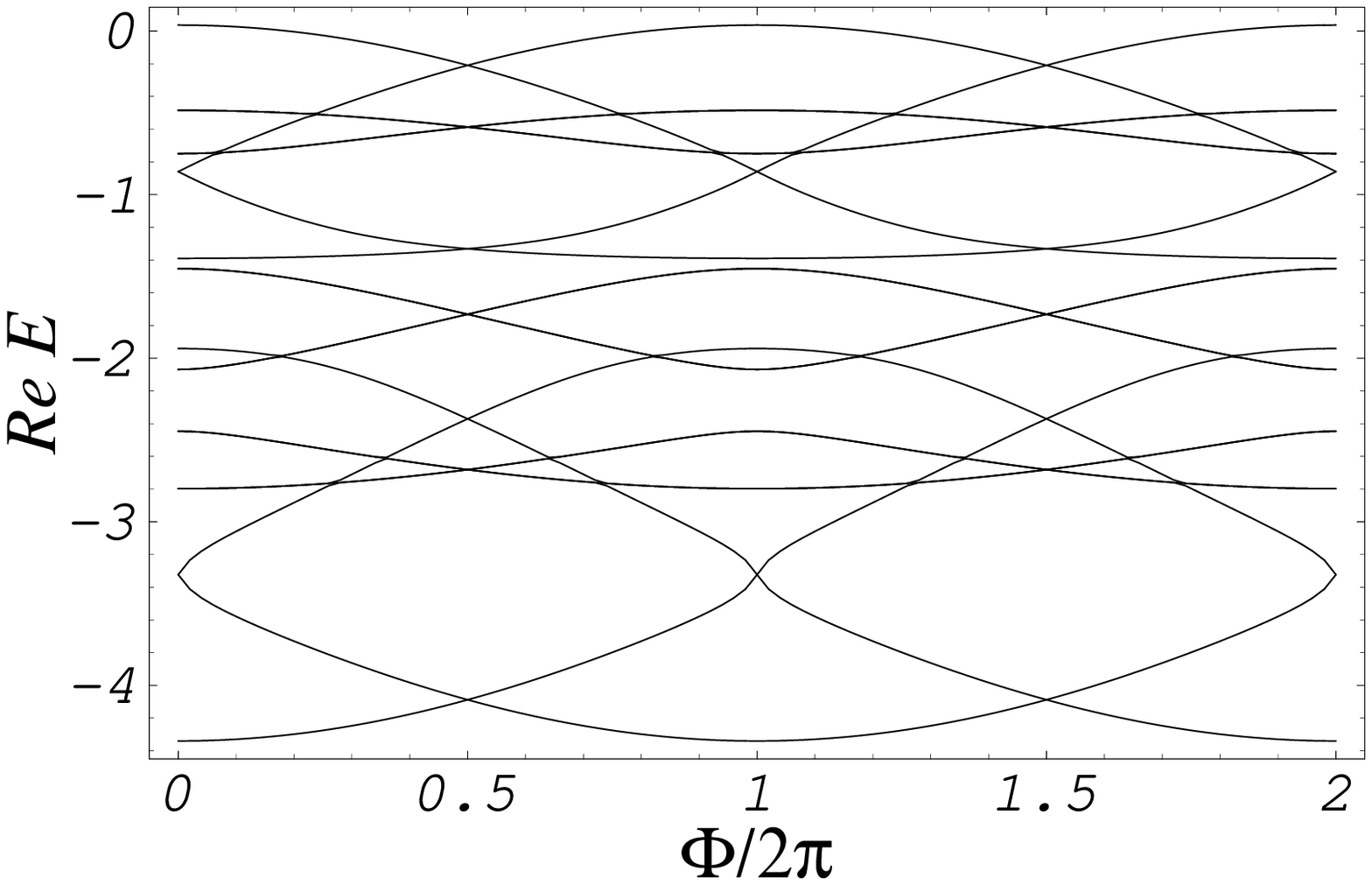}} 
\vspace{0.2cm}
\caption{The same spectral flow but with $\Psi=1.2$}
\label{f:diag6_12}
\end{figure}

\begin{figure}[htb]
\epsfxsize=70mm 
\centerline{\epsfbox{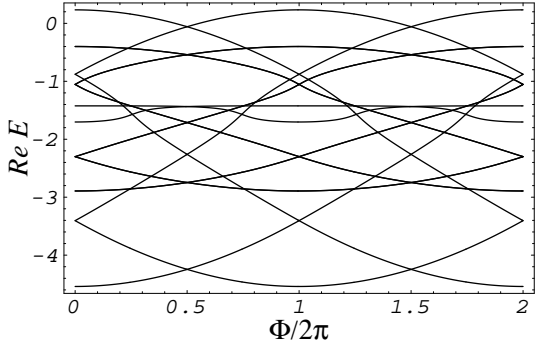}} 
\vspace{0.2cm}
\caption{The same spectral flow but with $\Psi=3$}
\label{f:diag6_30}
\end{figure}

We now extend the twist angle to a complex one. We should
hence  consider the
spectral flow on various paths on the complex $\Theta$ plane.
In order to see clearly how the spectral flow occurs, let us
regard the flow  as a function of $\Phi$ with a fixed $\Psi$.
For small $\Psi$, the flow is essentially the same as the $\Psi=0$ case,
as can be seen in Fig. \ref{f:diag6_11}, where only
the real part of $E$'s is plotted for simplicity.
Namely, except for the spectrum being complex, the
qualitative nature of the whole spectral flow is not modified by
a finite but sufficiently small $\Psi$.
This implies that such small $\Psi$ cannot 
collapse the $1/N$ gaps due to irrelevant backward scatterings.
It should be noted that at $\Phi=2\pi$ (mod $2\pi$) all eigenvalues
are real. If $\Psi$ is further increased,  the gap between 
the first and the 
second excited state becomes small and eventually disappears at a certain
$\Psi=\Psi_{\rm cr}$, above which these two states become
degenerate in the real part of $E$ (They are complex conjugate each 
other). Then the period of the ground state changes, as seen in
Fig. \ref{f:diag6_12}. In this case, the period of the ground state is
$8\pi$. If we further increase $\Psi$, the period becomes longer. For
example, in Fig. \ref{f:diag6_30}, it becomes macroscopic value
$2\pi N~ (=12 \pi)$.  Physically, the change of the period means that
the small $1/N$ gaps due to irrelevant interactions are 
successively closed by the non-hermitian effects, as mentioned above.
This phenomenon has the  essentially same origin as for the 
metal-insulator transition triggered by non-hermitian hoppings.

\begin{figure}[htb]
\epsfxsize=70mm 
\centerline{\epsfbox{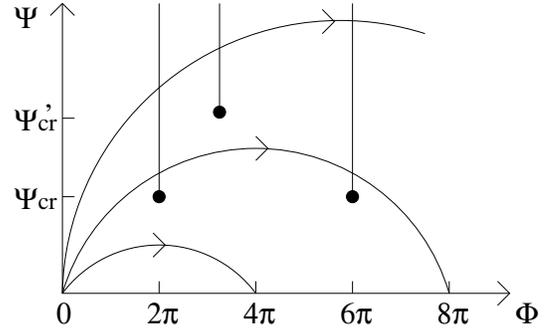}} 
\vspace{0.2cm}
\caption{Schematic illustration of the spectral flow of the ground
state starting at $\Theta=0$. The ground state follows the curves
towards the direction denoted by arrow and returns to the same ground
state for the first time at the end of the curves.}
\label{f:flow}
\end{figure}

Let us now consider the spectral flow defined 
on the whole complex $\Theta$ plane, which is sketched qualitatively
in Fig. \ref{f:flow}.  If we start with
 the ground state at $\Theta=0$, and trace a path in $\Theta$
plane, we may have  such spectral flows as  
depicted by several curves in this figure.
For small $\Psi$ the path can be deformed to the real axis without
changing the period. 
There exists, however,  certain critical $\Psi_{\rm cr}$
beyond which the gap between the first and the second excited states
disappear. We denote it as a straight line in the figure.
If we follow the path across this line, the period changes.
It should be noted that the structure of Fig. \ref{f:flow} is periodic
in $\Phi=2\pi$, though we have drawn only the lines relevant to the
flow of the ground state starting from $\Theta=0$.
Therefore, paths on the $\Theta$ plane are divided into several
classes classified in terms of the periods of the ground state.
Although the spectral flow on the complex plane has quite rich 
structure,  their essential property is 
understood  by studying the typical case  
where  the period of the ground state changes
from $4\pi$ to $8\pi$. 
In the following section, we will estimate 
the corresponding critical twist $\Psi_{\rm cr}$
based on the finite-size corrections.

\begin{figure}[htb]
\epsfxsize=70mm 
\centerline{\epsfbox{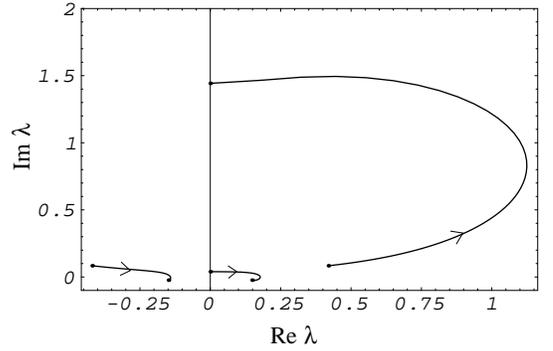}} 
\vspace{0.2cm}
\caption{Rapidity flow of the ground state of the $S^z=0$ sector with
$\Psi=1.1$, corresponding to Fig. \ref{f:diag6_11}. Initial ($\Phi=0$)
and final ($\Phi=2\pi$) points are denoted by dots, flowing towards
the direction indicated by arrows.}
\label{f:ba6_11}
\end{figure}

Let us now check explicitly how the corresponding rapidities flow.
The first example is Fig. \ref{f:ba6_11} 
which corresponds to Fig. \ref{f:diag6_11}.
\begin{figure}[htb]
\epsfxsize=70mm 
\centerline{\epsfbox{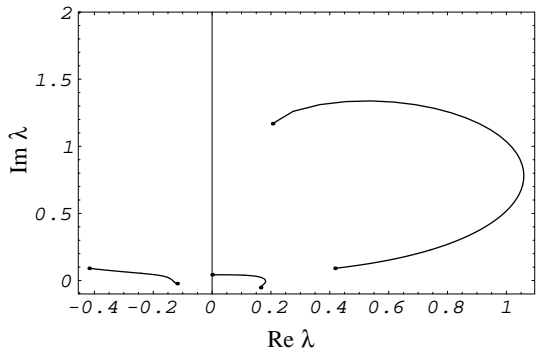}} 
\vspace{0.2cm}
\caption{The same rapidity flow but with $\Psi=1.2$,
corresponding to Fig. \ref{f:diag6_12}}
\label{f:ba6_12}
\end{figure}
\begin{figure}[htb]
\epsfxsize=70mm 
\centerline{\epsfbox{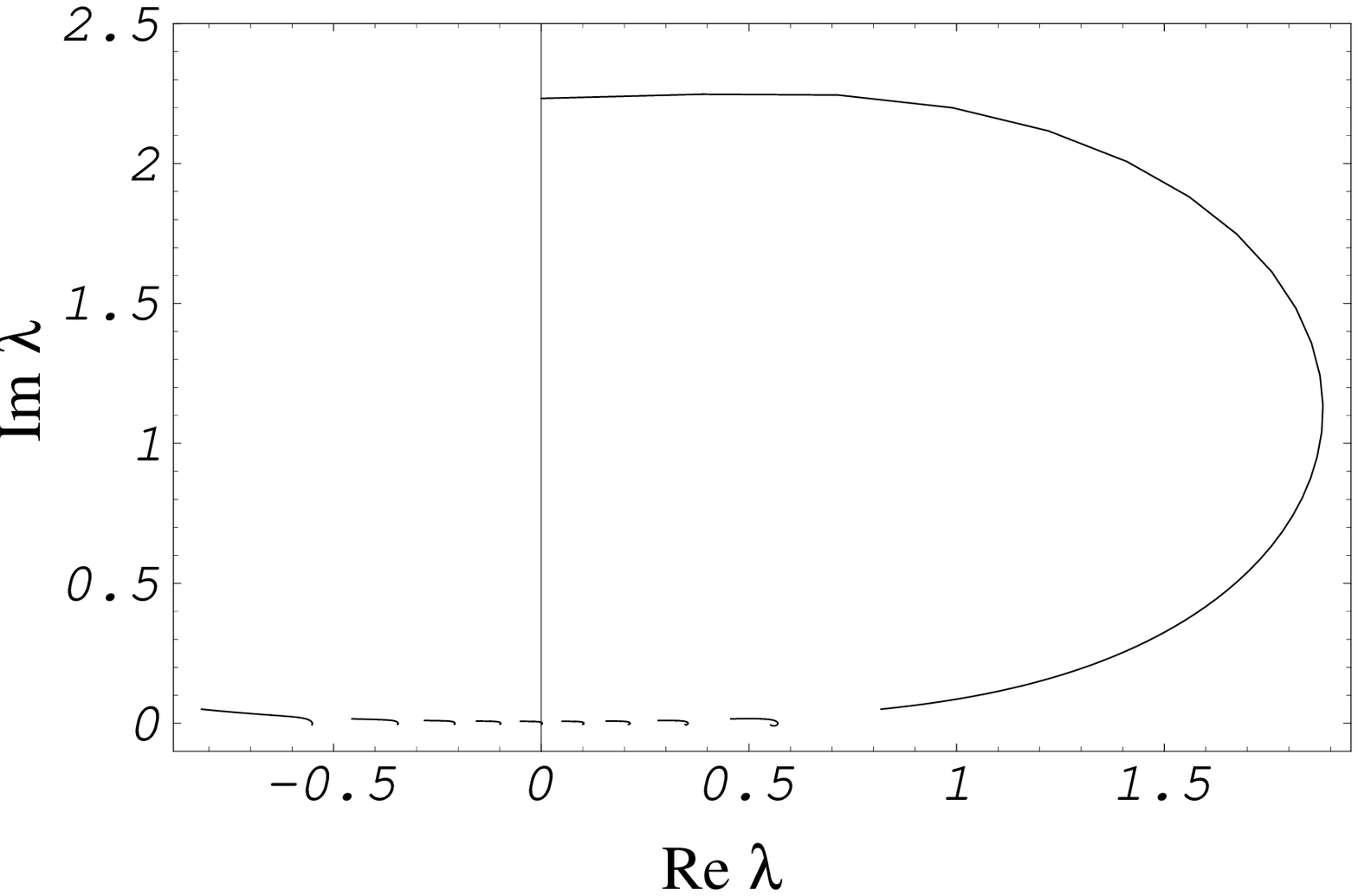}} 
\vspace{0.2cm}
\caption{The same rapidity flow but of $N=20$ system with $\Psi=0.7$}
\label{f:ba20_07}
\end{figure}
\begin{figure}[htb]
\epsfxsize=70mm 
\centerline{\epsfbox{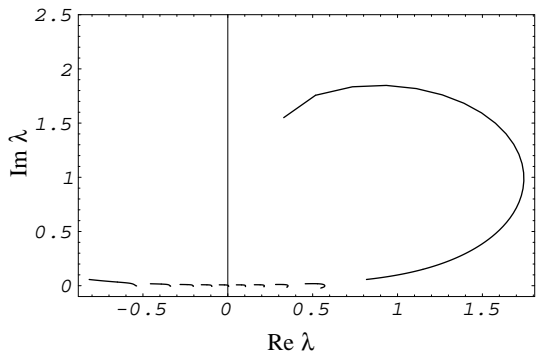}} 
\vspace{0.2cm}
\caption{The same rapidity flow but of $N=20$ system with $\Psi=0.8$}
\label{f:ba20_08}
\end{figure}
Since we are concerned with the $N=6$ site system, the ground state is 
described by three rapidities  sitting near 
the real axis at $\Phi=0$, indicated by
dots. They continuously move  on the complex plane, as $\Phi$ increases, 
towards the direction indicated by
arrows. Finally at $\Phi=2\pi$, rapidities reach the 
points indicated by dots. 
What should be noted here is the motion of the last rapidity:
After a long trip, it eventually reaches the imaginary axis. 
As a result, the whole distribution of the rapidities is changed
considerably, but it is still symmetric with
respect to the imaginary axis,  which
ensures that the total energy is real.
Contrary to this, for $\Psi=1.2>\Psi_{\rm cr}$ in Fig. \ref{f:ba6_12}
the final distribution
is not symmetric, which suggests that there exists another state
with the complex-conjugate energy corresponding to the configuration
reflected with respect to the imaginary axis. We have checked that
this is indeed true even for a larger system in Figs. \ref{f:ba20_07}
and \ref{f:ba20_08}.
Therefore, we can see that whether the symmetric solution exists 
at $\Phi=2\pi$ provides a  criterion whether the gap exits.
We have calculated $\Psi_{\rm cr}$ for several systems,
summarized in Table \ref{t:PsiCr}.
\begin{table}[htb] 
\begin{center}
\renewcommand{\arraystretch}{1.2}
\begin{tabular}{c|llllll}
$N$               &6     &8     &10    &20    &30    &50     \\
\hline
$\Psi_{\rm cr}$   &1.1456&1.0225&0.9446&0.7655&0.6899&0.6141\\
$\lambda_M/i$     &1.208 &1.308 &1.381 &1.622 &1.759 &1.915 \\
\hline
$\Psi_{\rm max}$  &1.0480&0.9532&0.8898&0.7363&0.6685&0.5991\\
$\lambda$         &1.372 &1.446 &1.507 &1.712 &1.837 &1.999\\
\hline
$\Psi_{N,{\rm max}}$ &1.1806&1.0536&0.9726&0.7861&0.7074&0.6287\\
$\lambda$         &1.323 &1.398 &1.461 &1.673 &1.802 &1.967
\end{tabular}
\end{center}
\caption{The values $\Psi_{\rm cr}$ calculated by several ways.
``$\Psi_{\rm cr}$'' is the exact value calculated by the Bethe ansatz
equations. ``$\Psi_{\rm max}$'' is the maximum value of the function 
$\Psi(\lambda)$ in (\ref{PsiFunExa}) by using the solution of 
Eq.(\ref{BAEReal}). See III. B. 
``$\Psi_{N,{\rm max}}$'' is the maximum value of the function 
$\Psi_N(\lambda)$ in (\ref{FinalPsi}).} 
\label{t:PsiCr}
\end{table}

By using the 
criterion mentioned above, we 
 wish to propose a practical method to  estimate the critical value 
$\Psi_{\rm cr}$ for a large but finite-size system. 
To this end let us assume that $\lambda_M$ sits on the
imaginary axis. Then if $\Psi<\Psi_{\rm cr}$, the Bethe ansatz
equations on this assumption have a solution, while for 
$\Psi>\Psi_{\rm cr}$, they do not have a solution.
Therefore, the critical value of $\Psi$ beyond which there exist 
such solutions can be identified with $\Psi_{\rm cr}$.
The problem becomes simpler if we put it in a different way:
We have so far solved the Bethe ansatz equations for a given $\Psi$.
We now set the problem for seeking for a possible 
$\Psi$ provided that the value of imaginary $\lambda_M$ is 
initially given. 
To solve this new problem, let us rewrite the Bethe ansatz equations at
$\Phi=2\pi$,
\begin{eqnarray}
&&Np(\lambda_j)+\sum_{k\ne j}^{M-1}\theta(\lambda_j-\lambda_k)
+\theta(i\lambda-\lambda_k)
\nonumber\\&&\qquad                                                 
=2\pi(I_j+1)+i\Psi(\lambda), 
\label{BAERealFull}\\
&&i\Psi(\lambda)=Np(i\lambda)+
\sum_{k=1}^{M-1}\theta(i\lambda-\lambda_k)-2\pi(I_M+1)
\label{PsiFunExa}
\end{eqnarray}
In the above equations we should  determine $\lambda_j$ 
($j=1,\cdots,M-1$) and $\Psi$ for a given real $\lambda$.
The equations do not have a solution for $\Psi>\Psi_{\rm cr}$ 
so that for any given $\lambda$, $\Psi(\lambda)$ should not
exceed $\Psi_{\rm cr}$. Namely, the function $\Psi(\lambda)$ should
have a maximum value, which in turn gives $\Psi_{\rm cr}$.
Therefore, the problem is now 
 to determine the function $\Psi(\lambda)$,
and find its maximum, which indeed allows us to obtain the critical
 twist $\Psi_{\rm cr}$. 

\subsection{Bethe ansatz equations in a decoupling approximation}

\begin{figure}[htb]
\epsfxsize=70mm 
\centerline{\epsfbox{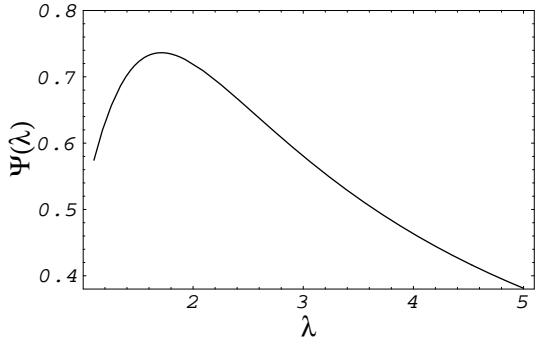}} 
\vspace{0.2cm}
\caption{Function $\Psi(\lambda)$ in eq.(\ref{PsiFunExa}) 
for $N=20$ system. 
We use the solutions of eq.(\ref{BAEReal})
as the nine rapidities $\lambda_k$ in this equation.}
\label{f:fssys20}
\end{figure}

Even if we employ the above trick, it is not so easy to
determine $\Psi_{\rm cr}$. 
We here propose an approximate treatment to the problem and confirm
that it works fairly well for $N=20$ system.
Suppose that $N$ and $\lambda$ are large. Then the solutions of 
eq.(\ref{BAEReal}) can be regarded as approximate solutions.
Therefore, if we substitute them into eq.(\ref{PsiFunExa}) and plot 
the result as a function of $\lambda$, extrapolating 
to the small $\lambda$ region, we have, for example,
Fig. \ref{f:fssys20}. This function indeed has a maximum value, which
explains the existence of $\Psi_{\rm cr}$. 
It is remarkable that the maximum value thus obtained 
is quite close to the exact one in Table 
\ref{t:PsiCr}. We list the maximum values of $\Psi(\lambda)$ and 
corresponding $\lambda$ in the same Table. 
They deviate from the exact value by only a few
percent in general. Therefore, we can say that 
the value of $\Psi_{\rm cr}$ is essentially determined by 
the rapidity $\lambda_M$ on the imaginary axis, and
our approximate treatment works quite well.

In order to calculate $\Psi_{\rm cr}$ self-consistently,
we have to calculate the finite-size corrections based on 
eqs.(\ref{BAERealFull}) and (\ref{PsiFunExa}).
There are various $O(1/N)$ terms, all of which should be taken into 
account. As we have already checked for the $N=20$ system, 
however, the contributions from the third term of the
left-hand side and the second term of the right hand side in 
eq.(\ref{BAERealFull}), i.e., $\theta(i\lambda-\lambda_k)$ and 
$i\Psi(\lambda)$, respectively, are quite small.
Therefore, we set 
$\theta(i\lambda-\lambda_k)\rightarrow\pi$ and 
$i\Psi(\lambda)\rightarrow0$ in this equation, and  
confirm the validity of this treatment 
by comparing the obtained results on this assumption with 
the exact $\Psi_{\rm cr}$ for a small system.
Thus, the Bethe equations for
$\lambda_j$ ($j=1,\cdots,M-1$) are decoupled from that 
for $i\lambda\equiv\lambda_M$, so that we use the notations $\mu_j$
for these $\lambda_j$'s. 
In order to proceed the calculation,
let us introduce the following functions from eqs.(\ref{BAEReal}) and
(\ref{PsiFun})
\begin{eqnarray}
&&Z_N(\mu)=\frac{p(\mu)}{2\pi}+\frac{1}{2\pi N}
\sum_{k}\theta(\mu-\mu_k),
\nonumber\\
&&i\Psi_N(\lambda)=Np(i\lambda)+\sum_k\theta(i\lambda-\mu_k)
-\pi\left(\frac{N}{2}+1\right),
\label{PsiFun}
\end{eqnarray}
where $Z_N(\mu_k)=I_k'/N$.
All $\mu$'s are on the real axis.
The distribution function is given by
\begin{equation}
\sigma_N(\mu)=\frac{p'(\mu)}{2\pi}+\frac{1}{2\pi N}
\sum_{k}\theta'(\mu-\mu_k).
\label{DecDis}
\end{equation}
By definition,
\begin{equation}
\int_{-\infty}^\infty d\mu\sigma_N(\mu)=\frac{1}{2}+\frac{1}{N}.
\end{equation}
The above equations enable us to  further proceed  the  
analytic calculation, which will be performed separately 
in the next section.

\section{Finite-size corrections}

We are now ready to evaluate finite-size corrections
based on the equations in the preceding section. In particular, 
we derive the $\Psi_{\rm cr}$ as a maximum value of 
the function $\Psi(\lambda)$
in the bulk limit with leading finite-size corrections.
So far, practical methods to compute the finite-size
corrections for non-hermitian models 
\cite{Kim,NohKim} have not been well explored.
So, we hope that the following treatment may also
shed some light on such problems.

\subsection{Bulk limit}

We wish to first investigate  Eqs.(\ref{DecDis}) and (\ref{PsiFun})
in the bulk limit $N\rightarrow\infty$ 
to confirm that our approach is indeed consistent.
Since the effects of the twist angle should be $O(1/N)$, 
we expect that the function $\Psi(\lambda)$ in the bulk
limit is identically zero for any $\lambda$, which
is shown to be indeed satisfied in our treatment.
In the bulk limit, Eqs.(\ref{DecDis}) and (\ref{PsiFun}) are 
\begin{eqnarray}
&&\sigma_\infty(\mu)=\frac{p'(\mu)}{2\pi}+\frac{1}{2\pi}
\int_{-\infty}^\infty d\mu'
\theta'(\mu-\mu')\sigma_\infty(\mu'),
\label{BulDis}\\
&&\frac{i\Psi_\infty(\lambda)}{N}=
p(i\lambda)+\int_{-\infty}^\infty d\mu
\theta(i\lambda-\mu)\sigma_\infty(\mu)
-\frac{\pi}{2},
\label{BulPsi}
\end{eqnarray}
where the left-hand side of the second equation means 
$\lim_{N\rightarrow\infty}\frac{i\Psi_N(\lambda)}{N}$.
Since $\mu$'s are distributed on the real axis,
their distribution can be 
easily determined by the Fourier transformation.
The result is  
\begin{equation}
\sigma_\infty(\mu)=\frac{1}{2}\sech\pi\mu.
\end{equation}
By substituting this solution into eq.(\ref{BulPsi}), we have
\begin{eqnarray}
\frac{i\Psi_\infty(\lambda)}{N}=&&
i\int_{-\infty}^\infty d\mu
\frac{\sech\pi\mu}{2}
\ln\left(\frac{\mu-i(\lambda-1)}{\mu-i(\lambda+1)}\right)
\nonumber\\&&                                                      
-i\ln\left(\frac{\lambda-1/2}{\lambda+1/2}\right),
\end{eqnarray}
where we have assumed $\lambda>1$.
The first term of the right-hand side is evaluated by extending 
the contour integral in the lower half plane and 
summing up the infinite series 
of residues on the imaginary axis. We find that it cancels the
second term, and leads to 
\begin{equation}
\frac{i\Psi_\infty(\lambda)}{N}=0.
\end{equation}
This proves that our treatment indeed produces 
the correct result in the bulk limit
because effects of the twist should disappear in this limit.

\subsection{$1/N$ corrections}

So far we have checked that $\Psi_{\infty}(\lambda)=0$
for any $\lambda>1$, 
which means that the gap in the spectrum always disappears
in the bulk limit. It is quite natural because the 
gap is caused by marginally irrelevant interaction.
Now we proceed to finite-size systems, for which 
we expect a finite $\Psi_{\rm cr}$.

Subtracting eq.(\ref{BulDis}) from eq.(\ref{DecDis})
we have the identity valid for a system with arbitrary $N$,
\begin{eqnarray}
\sigma_{N}(\mu)-\sigma_{\infty}(\mu)=&&
-\int_{-\infty}^\infty d\mu'q(\mu-\mu')
\nonumber\\&&\times                                                
\left[\frac{1}{N}\sum_k\delta(\mu'-\mu_k)
-\sigma_N(\mu')\right],
\label{FinDis}
\end{eqnarray}
where the kernel $q(\mu)$ is
\begin{equation}
q(\mu)=\frac{1}{2\pi}\int_{-\infty}^\infty d\omega e^{-i\omega\mu}
\frac{1}{e^{|\omega|}+1}.
\end{equation}
Denote the largest rapidity on the real axis as $\Lambda$.
Then $Z_N(\Lambda)=\frac{1}{4}-\frac{1}{N}$, or equivalently,
\begin{equation}
\int_{-\Lambda}^\Lambda d\mu\sigma_N(\mu)=\frac{1}{2}-\frac{2}{N}.
\end{equation}
Similarly, $\Psi(\lambda)$ is rewritten as 
\begin{eqnarray}
\frac{i\Psi_N(\lambda)}{N}=&&
\int_{-\infty}^\infty d\mu Q(\lambda,\mu)
\nonumber\\&&\times                                                
\left[\frac{1}{N}\sum_k\delta(\mu-\mu_k)-\sigma_N(\mu)\right]
-\frac{\pi}{N},
\label{FinPsiD}
\end{eqnarray}
where
\begin{equation}
Q(\lambda,\mu)
=\theta(i\lambda-\mu)-\int_{-\infty}^\infty d\mu'
\theta(i\lambda-\mu')q(\mu'-\mu),
\end{equation}
The kernel $Q(\lambda,\mu)$ can be evaluated by counter integral in
the lower half plain, yielding the result
\begin{eqnarray}
&&Q(\lambda,\mu)=-\frac{\pi}{2}+\widetilde Q(\lambda,\mu), 
\nonumber\\
&&\widetilde Q(\lambda,\mu)=
i\ln\left(\frac{\mu-i(\lambda-1)}{\mu-i\lambda}\right).
\label{KerQ}
\end{eqnarray}
Substituting this equation into Eq.(\ref{FinPsiD}), 
we find that the first constant
term in Eq.(\ref{KerQ})
cancels the last constant term in Eq.(\ref{FinPsiD}). 
Therefore,
\begin{eqnarray}
\frac{i\Psi_N(\lambda)}{N}=&&
\int_{-\infty}^\infty d\mu\widetilde Q(\lambda,\mu)
\nonumber\\&&\times                                                
\left[\frac{1}{N}\sum_k\delta(\mu-\mu_k)-\sigma_N(\mu)\right].
\label{FinPsi}
\end{eqnarray}

To evaluate the leading finite-size effects, let us 
apply the Euler-Maclaurin formula to 
Eqs.(\ref{FinDis}) and (\ref{FinPsiD}). Up to $O(N^{-2})$, we have
\begin{eqnarray}
&&\sigma_N(\mu)-\sigma_\infty(\mu)
=\int_\Lambda^\infty d\mu'q(\mu-\mu')\sigma_N(\mu')
\nonumber\\&&\qquad                                                 
-\frac{q(\mu-\Lambda)}{2N}
+\frac{q'(\mu-\Lambda)}{12N^2\sigma_N(\Lambda)}
+(-\Lambda ~{\rm terms}),
\label{EMsig}\\
&&\frac{i\Psi_N(\lambda)}{N}=
-2\int_\Lambda^\infty d\mu\widetilde Q(\lambda,\mu)\sigma_N(\mu)
\nonumber\\&&\qquad                                                 
+\frac{\widetilde Q(\lambda,\Lambda)}{N}
+\frac{\widetilde Q'(\lambda,\Lambda)}{6N^2\sigma_N(\Lambda)},
\label{EMPsi}
\end{eqnarray}
where $\widetilde Q'(\lambda,\Lambda)=
\partial\widetilde Q(\lambda,\Lambda)/\partial\Lambda$.
The cut-off $\Lambda$ is determined by the condition
\begin{equation}
\int_\Lambda^\infty d\mu\sigma_N(\mu)=\frac{3}{2N}.
\label{DetLam}
\end{equation}

\subsubsection{Finite size corrections to $\sigma_N(\mu)$}

As we have exploited the decoupling approximation, we can evaluate the
finite-size corrections to $\sigma_N(\lambda)$ 
following  standard techniques \cite{VegWoy,WoyEck,WET,HQB}.
Introducing
\begin{eqnarray}
&&\sigma_N^+(\mu)=\left\{
\begin{array}{ll}\sigma_N(\mu+\Lambda) & \quad {\rm for}\quad\mu>0\\
0 & \quad {\rm for}\quad\mu<0,
\end{array}\right.
\nonumber\\
&&\sigma_N^-(\mu)=\left\{
\begin{array}{ll}0 & \quad {\rm for}\quad\mu>0\\
\sigma_N(\mu+\Lambda) & \quad {\rm for}\quad\mu<0,
\end{array}\right. 
\end{eqnarray}
we can write down eq.(\ref{EMsig}) in the region $\mu\sim\Lambda$,
by neglecting $-\Lambda$ terms,
\begin{eqnarray}
\sigma_N^+(\mu)&&+\sigma_N^-(\mu)
-\int_{-\infty}^\infty d\mu'q(\mu-\mu')\sigma_N^+(\mu')
\nonumber\\&&                                                        
=\sigma_\infty(\mu+\Lambda)-
\frac{q(\mu)}{2N}+\frac{q'(\mu)}{12N^2\sigma_N(\Lambda)} .
\label{WieHop}
\end{eqnarray}
Solving the equation in a standard way summarized briefly in the 
appendix, we end up with the final results
\begin{equation}
\widetilde\sigma_N^+(\omega)=
C(\omega)+G_+(\omega/2\pi)\left[P(\omega)+Q_+(\omega)\right],
\end{equation}
where $\widetilde\sigma_N^\pm(\omega)$ 
is the Fourier-transformation of $\sigma_N^\pm(\lambda)$ and where
\begin{eqnarray}
&&C(\omega)=\frac{1}{\sqrt{2\pi}}
\left(\frac{1}{2\pi}+\frac{i\omega}{12N^2\sigma_N(\Lambda)}\right) ,
\nonumber\\
&&G_+(\omega/2\pi)=
\sqrt{2\pi}\left(-\frac{i\omega}{2\pi e}\right)^{-\frac{i\omega}{2\pi}}
/\Gamma\left(\frac{1}{2}-\frac{i\omega}{2\pi}\right),
\nonumber\\
&&P(\omega)=-\frac{1}{\sqrt{2\pi}}
\left(\frac{1}{2N}-\frac{2\pi ig}{12N^2\sigma_N(\Lambda)}
+\frac{i\omega}{12N^2\sigma_N(\Lambda)}\right),
\nonumber\\
&&Q_+(\omega)=
\frac{i}{\sqrt{2\pi}}
\frac{e^{-\pi\Lambda}}{\omega+\pi i}
G_+\left(i/2\right),
\end{eqnarray}
with 
\begin{eqnarray}
&&ig=1/24,
\nonumber\\
&&\sigma_N(\Lambda)=\frac{\pi}{N}\alpha,
\nonumber\\
&&\Lambda=\frac{1}{\pi}\ln\left(\frac{N}{\sqrt{\pi e}\beta}\right).
\end{eqnarray}
Here several constants are given by
\begin{eqnarray}
&&\alpha=\frac{1}{2}\left(a_0+\sqrt{a_0^2-4a_1}\right), 
\nonumber\\
&&\beta=\frac{1}{2}+\frac{1}{\sqrt{2}}-\frac{ig}{6\alpha},
\nonumber\\
&&a_0=\frac{1}{2}+\frac{1}{\sqrt{2}}+ig,
\nonumber\\
&&a_1=\frac{ig+(ig)^2}{6}.
\end{eqnarray}

\subsubsection{Finite-size corrections to $\Psi_N(\lambda)$}

By using the preceding results for $\sigma_N(\lambda)$, 
let us now calculate $\Psi_N(\lambda)$ with finite-size effects.
To begin with, we convert the first term in the right-hand side of
eq.(\ref{EMPsi}) into the integral in Fourier space,
\begin{eqnarray}
&&-2\int_\Lambda^\infty d\mu\widetilde Q(\lambda,\mu)\sigma_N(\mu)
\nonumber\\&&                                                      
=-i\sqrt{2\pi}\int_{-\infty}^\infty d\omega e^{i\Lambda\omega}
\frac{e^{-\lambda|\omega|}-e^{-(\lambda-1)|\omega|}}{|\omega|}
\widetilde\sigma_N^+(\omega).
\end{eqnarray}
It is easily shown that $C(\omega)$ in $\widetilde\sigma_N^+(\omega)$
cancels out the second and third terms in
 the right-hand side of eq.(\ref{EMPsi}).
Therefore, we eventually arrive at 
\begin{eqnarray}
\Psi_N(\lambda)=&&
-\int_{-\infty}^\infty d\omega e^{i\Lambda\omega}
\frac{e^{-\lambda|\omega|}-e^{-(\lambda-1)|\omega|}}{|\omega|}
\nonumber\\&&\times                                                
G_+(\omega/2\pi)(P(\omega)+Q_+(\omega)).
\label{FinalPsi}
\end{eqnarray}

\begin{figure}[htb]
\epsfxsize=70mm 
\centerline{\epsfbox{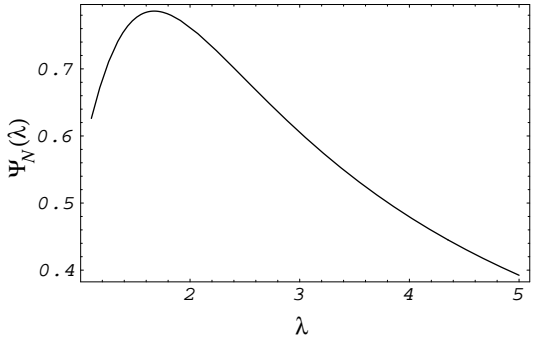}} 
\vspace{0.2cm}
\caption{$\Psi_N(\lambda)$ for $N=20$ system}
\label{f:fsc20}
\end{figure}

We have evaluated the integral numerically for the $N=20$ system 
in Fig. \ref{f:fsc20}, corresponding
to Fig. \ref{f:fssys20}.
As we expect, the leading finite-size calculation yields
the desirable results.
In fact, the maximum value of this function reproduces 
the critical $\Psi$ expected for 
small systems fairly well, which are listed in Table \ref{t:PsiCr}.
\begin{figure}[htb]
\epsfxsize=70mm 
\centerline{\epsfbox{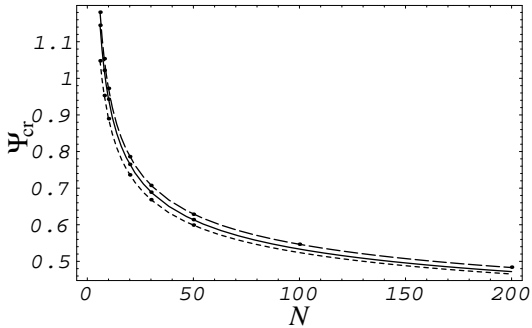}} 
\vspace{0.2cm}
\caption{The data of $\Psi_{\rm cr}$ in Table \ref{t:PsiCr} and
their least-square fits. See the text.}
\label{f:compa}
\end{figure}

It seems not easy to precisely determine the large-$N$ behavior of the
function $\Psi(\lambda)$.  However, 
observing the data for several small $N$ systems 
of the exact Bethe ansatz
calculation and also for larger $N$ systems of Eq.(\ref{FinalPsi}),
 we can predict $1/\ln N$ behavior in the asymptotic
regime. To confirm this explicitly, we plot in Fig. \ref{f:compa}
the least-square fit using $(\ln N)^{-n}$ with $n=1,2,3$
with the data points.
The solid-line is for the exact Bethe ansatz results, while the
dotted- and dashed-lines are for the decoupled 
Bethe ansatz results of 
finite-size systems in section III. B.
and for Eq.(\ref{FinalPsi}), respectively.
As expected,  the larger the system is,
the smaller the difference between the dotted- and dashed-lines
becomes.
Furthermore, these $\Psi_{\rm cr}$ in a decoupling approximation are
very close to the exact $\Psi_{\rm cr}$.
Note that $\Psi_{\rm cr}$ is not a gap itself between
the first and second excited states. The gap of $O(1/N)$ exist
there, and we need $\Psi_{\rm cr}$ of $O(1/\ln N)$ to close this
gap.  

\section{Summary and Discussions}

Stimulated by the recent advance in the 
theory for localization-delocalization
transitions in 1D non-hermitian systems, 
we have studied how the non-hermitian property affects the integrable
Heisenberg spin model. In particular, we have focused the
particular attention on the spectral flow by imposing
twisted boundary conditions with  complex twist.
By solving the Bethe ansatz equations numerically for finite systems,
we have shown that 
a global structure of the spectral flow completely changes
its character as the imaginary twist increases. Namely, the period 
of the ground state increases from $4\pi$  
to $8\pi $ at a certain imaginary twist angle.
If the imaginary twist is further increased, the period successively
jumps to a higher value at the corresponding imaginary twist angle.
The origin of this phenomenon has been clarified:
the small $1/N$ energy gaps for a finite system caused by 
irrelevant interactions are
closed by the non-hermitian hopping of XY term.
To determine the critical imaginary twist for the  transition, 
we have also proposed a practical method to evaluate 
the  finite-size corrections, 
and discussed the behavior in the large $N$ regime. 
Although we have neglected some terms which may 
give $O(1/N)$ contributions
in this treatment, the estimated value of $\Psi_{\rm cr}$ is quite
close to the exact one. Since practical ways to 
explicitly evaluate the finite-size corrections for non-hermitian
models have not been developed well, the present treatment 
may provide a guideline to study such problems in more detail.

In this paper we have treated $\Phi$ and $\Psi$ on an equal footing. 
Before concluding the paper, 
we briefly make a comment on the physical origin of
these twists. As mentioned in the introduction, 
the real twist $\Phi$ can be 
naturally regarded as an Aharonov-Bohm flux. As for imaginary 
$\Psi$, its origin may be ascribed to randomness in the system.
This is indeed seen by observing  the Hamiltonian,
\begin{equation}
H=\sum_j\left[
\frac{1}{2}\left(
e^{-\Psi_j}S_j^+S_{j+1}^-+e^{\Psi_j}S_j^-S_{j+1}^+
\right)
+S_j^zS_{j+1}^z
\right]
\label{HamRan}
\end{equation}
with the boundary condition
\begin{equation}
S_{j+N}^{\pm}=e^{\pm i\Phi}S_j,\quad S_{j+N}^z=S_j^z.
\end{equation}
where $\Psi_j$ are not necessarily uniform.
The Bethe ansatz equation is the same as Eq.(\ref{BAE}),
but with
\begin{equation}
\Theta=\Phi+i\sum_j\Psi_j.
\end{equation}
Therefore, it is seen that our imaginary twist can 
be regarded as the consequence of 
non-hermitian random hoppings.
By taking into account the ensemble average over random
variables $\Psi_j$, we may have a chance to discuss the effects of
random non-hermitian hoppings, which should be 
treated in the future work. 

The Hamiltonian (\ref{HamRan}) is also useful to see how the 
present analysis is related to 
the metal-insulator transition in the asymmetric models, and also
to the delocalization transition in the Hatano-Nelson model.
To this end, let us set $\Psi_j=\Psi/N$, 
where $\Psi$ is given by (\ref{TwiAng}). 
By using the hard-core boson representation,
the XY term is converted into the hopping term
of bosons, $H_{\rm hop}=\sum_j(e^{-\Psi/N}b_j^\dagger b_{j+1}
+e^{\Psi/N}b_{j+1}^\dagger b_j)$, where $b_j$ is a boson operator.
Namely, the present model is equivalent to the interacting 
bosonic lattice model with an imaginary vector potential. Note that the 
kinetic part is essentially the same as that for 
the Hatano-Nelson model.
For the latter model, Hatano and Nelson demonstrated that the ``insulating'' 
state caused by randomness can be delocalized and 
become metallic at a certain critical $\Psi$ when the strength of 
the imaginary vector potential is increased.
This implies that the transition is caused by the 
competition between the 
imaginary vector potential and the random potential.
We can also observe a similar insulator-metal transition 
caused by imaginary vector potential
for the periodic model such as the asymmetric XXZ model
with Ising-type anisotropy\cite{ADW}, for which 
the massive (insulating) state is formed by the many-body interaction,
instead of randomness. In contrast  to the above examples, for 
the isotropic Heisenberg model (\ref{HamRan}) we have dealt with, 
the system is massless and ``metallic'' in the thermodynamic limit
even for $\Psi=0$. However, we should recall that
there still exist small (mesoscopic) gaps formed 
by irrelevant many-body interactions for a {\it finite-size system}, 
which control the period of the spectral flow for the ground state.
We have shown explicitly via the present study
that the competition between the imaginary vector potential and the
many-body interaction indeed modifies the period, which implies that
the mesoscopic gaps are closed by  the 
imaginary vector potential: Our transition is characterized by the 
collapse of the mesoscopic gaps, whereas that for the
XXZ case is by the  collapse of the ordinary gap.
We can now see that there are deep relationship among
the transition found in the spectral flow of the present model, 
the ``insulator-metal''
transition for the XXZ model, and the Hatano-Nelson model, 
although at first sight our results may not have any essential 
connections to latter phenomena;  
they may indeed share some essential features, e.g., the transition 
is driven by imaginary vector potentials and is
characterized by complex eigenvalues at criticality.
It may be interesting to study wider class of models with imaginary
vector potentials to understand a novel class of the
metal-insulator transition in a unified framework.

\acknowledgements
The authors would like to thank H.-P. Eckle, A. Kl\"umper
and M. Chiba for valuable discussions.
Numerical computation in this work was carried out at the 
Yukawa Institute Computer Facility.
This work is partly
supported by the Grant-in-Aid from the Ministry of
Education, Science and Culture, Japan.

\appendix
\section{Wiener-Hopf equation}

Since there exist extensive studies of finite-size 
calculations\cite{VegWoy,WoyEck,WET,HQB}, we
summarize it briefly for reference sake.

Introducing the Fourier transformation with the normalization
\begin{equation}
\sigma_N^\pm(\lambda)=\frac{1}{\sqrt{2\pi}}\int_{-\infty}^\infty d\omega
e^{-i\lambda\omega}\widetilde\sigma_N^\pm(\omega),
\end{equation}
Eq.(\ref{WieHop}) is given by the Fourier component
\begin{equation}
\widetilde\sigma_N^-(\omega)+
\frac{\widetilde\sigma_N^+(\omega)-C(\omega)}{e^{-|\omega|}+1}
=\frac{e^{-i\omega\Lambda}\sech\frac{\omega}{2}}{2\sqrt{2\pi}}-C(\omega)
\label{WieHopFou}
\end{equation}
The kernel is now factorized as
\begin{equation}
e^{-|\omega|}+1=G_+(\omega/2\pi)G_-(\omega/2\pi),
\end{equation}
where $G_\pm(z)$ is holomorphic in the upper-(lower) half plain,
respectively. 
The asymptotic form is
\begin{equation}
G_+(z)=1+\frac{g}{z}+\frac{g^2}{2z^2}+O(z^{-3}).
\label{AsyG}
\end{equation}
Then Eq.(\ref{WieHopFou}) can be written as
\begin{eqnarray}
&&\frac{\widetilde\sigma_N^+(\omega)-C(\omega)}{G_+(\omega/2\pi)}
-Q_+(\omega)
\nonumber\\&&\quad                                                  
=-G_-(\omega/2\pi)(\widetilde\sigma_N^-(\omega)+C(\omega))+Q_-(\omega)
\nonumber\\&&\quad                                                  
\equiv P(\omega).
\end{eqnarray}
where 
$Q_\pm(\omega)$ is holomorphic in the upper-(lower) half plain,
satisfying
\begin{equation}
G_-(\omega/2\pi)
\frac{e^{-i\omega\Lambda}\sech\frac{\omega}{2}}{2\sqrt{2\pi}}
=Q_+(\omega)+Q_-(\omega),
\end{equation}
where we take into account only the contribution from the 
pole nearest to the real 
axis in the text. From the asymptotic 
form Eq.(\ref{AsyG}), $P(\omega)$ is determined,
which is summarized in the text.

The parameters $\Lambda$ and $\sigma(\Lambda)$ are determined by the
conditions Eq.(\ref{DetLam}) and
\begin{equation}
\sigma_N(\Lambda)=\frac{2}{\sqrt{2\pi}}\int_{-\infty}^{\infty}d\omega
\widetilde\sigma_N^+(\omega),
\end{equation}
or more precisely,
\begin{eqnarray}
&&G_+(i/2)e^{-\pi\Lambda}=\left(\frac{1}{2}+\frac{1}{\sqrt{2}}
\right)\frac{\pi}{N}-\frac{\pi (2\pi ig)}{12N^2\sigma_N(\Lambda)},
\nonumber\\
&&\sigma_N(\Lambda)=G_+(i/2)e^{-\pi\Lambda}+\frac{(2\pi ig)}{2N}
-\frac{(2\pi ig)^2}{24N^2\sigma_N(\Lambda)}.
\end{eqnarray}
The solutions are summarized in the text.



\begin{references}
\bibitem[*]{Email} Email: fukui@yukawa.kyoto-u.ac.jp
\bibitem{Veg} H. J. de Vega, 
Nucl. Phys. {\bf B240} (1984) 495.
%
\bibitem{ABB} F. C. Alcaraz, M. N. Barber and M. T. Batchelar, 
Ann. Phys. {\bf 182} (1988) 280.
%
\bibitem{YunBat} C. M. Yung and M. T. Batcherlor, 
Nucl. Phys. {\bf B446} (1995) 461.
%
\bibitem{KBI} V.E. Korepin, N.M. Bogoliubov and A.G. Izergin,
``{\it Quantum Inverse Scattering Method and Correlation Functions}'',
Cambridge University Press (1993).
%
\bibitem{ByeYan} N. Byers and C. N. Yang,
Phys. Rev. Lett. {\bf 7} (1961) 46.
\bibitem{Koh} W. Kohn, 
Phys. Rev. {\bf 133} (1964) 171.
%
\bibitem{ShaSut} B.S. Shastry and B. Sutherland, 
Phys. Rev. Lett. {\bf 65} (1990) 243.
\bibitem{SutSha} B. Sutherland and B. S. Shastry, 
Phys. Rev. Lett. {\bf 65} (1990) 1833.
%
\bibitem{YuFow} N. Yu and M. Fowler, 
Phys. Rev. {\bf 46} (1992) 14583.
%
\bibitem{RES} R.A. R\"omer, H.-P. Eckle and B. Sutherland,
Phys. Rev. {\bf B52} (1995) 1656.
\bibitem{Sut} B. Sutherland, 
Phys. Rev. Lett. {\bf 74} (1995) 816.
%
\bibitem{KusAok} K. Kusakabe and H. Aoki, 
J. Phys. Soc. Jpn {\bf 65} (1996) 2772. 
\bibitem{KorWu} V. E. Korepin and A. C. T. Wu, 
Int. J. Mod. Phys. {\bf B5} (1991) 497.
%
\bibitem{FIO} N. Fumita, H. Itoyama and T. Oota, 
Int. J. Mod. Phys. {\bf A12} (1997) 801.
\bibitem{FukKaw1}T. Fukui and N. Kawakami,
J. Phys. Soc. Jpn {\bf 65} (1996) 2824.
%
\bibitem{FukKaw2} T. Fukui and N. Kawakami, 
Phys. Rev. Lett. {\bf 76} (1996) 4242;
Phys. Rev. {\bf B54} (1996) 5346;
Nucl. Phys. {\bf B483} (1997) 663.
%
\bibitem{LiuWan} J. T. Liu and D. F. Wang, 
Phys. Rev. {\bf B55} (1997) R3344;
Phys. Rev. {\bf B56} (1997) 2312.
%
\bibitem{Hal} F. D. M. Haldane, 
Phys. Rev. Lett. {\bf 67} (1991) 937.
\bibitem{Kit} A. Kitazawa, 
J. Phys. {\bf A30} (1997) L285.
%
\bibitem{NasTir} T. Nassar and O. Tirkkonen,
hep-th/9707098.
\bibitem{HatNel} N. Hatano and D. R. Nelson, 
Phys. Rev. Lett. {\bf 77} (1997) 570;
cond-mat/9705290.
\bibitem{SYY} C. P. Yang, 
Phys. Rev. Lett. {\bf 19} (1967) 586;
B. Sutherland, C. N. Yang and C. P. Yang, 
Phys. Rev. Lett. {\bf 19} (1967) 588.
%
\bibitem{Nol} I. M. Nolden,
J. Stat. Phys. {\bf 67} (1992) 155.
%
\bibitem{BukSho} D. J. Bukman and J. D. Shore,
J. Stat. Phys. {\bf 78} (1995) 1277.
%
\bibitem{GwaSpo} L. H. Gwa and H. Spohn,
Phys. Rev. {\bf A46} (1993) 844.
%
\bibitem{ADHR} F. C. Alcaraz, H. Droz, M. Henkel and V. Rittenberg,
Ann. Phys. {\bf 230} (1994) 250.
%
\bibitem{NeeNij} J. Neergaard and M. den Nijs,
Phys. Rev. Lett. {\bf 74} (1995) 730.
%
\bibitem{Kim} D. Kim,
Phys. Rev. {\bf E52} (1995) 3512.
%
\bibitem{ADW} G. Albertini, S. R. Dahmen and B. Wehefritz, 
J. Phys. {\bf A29} (1996) L369; 
Nucl. Phys. {\bf B493} (1997) 541.
%
\bibitem{NohKim} J. D. Noh and D. Kim, cond-mat/9511001.
\bibitem{VegWoy} H. J. de Vega and F. Woynarovich,
Nucl. Phys. {\bf B251} (1985) 439.
%
\bibitem{WoyEck}  F. Woynarovich and H.-P. Eckle,
J. Phys. {\bf A20} (1987) L97.
%
\bibitem{WET} F. Woynarovich, H.-P. Eckle and T. T. Truong,
J. Phys. {\bf A22} (1989) 4027.
%
\bibitem{HQB} C. J. Hamer, G. R. W. Quispel and M. T. Batchelor,
J. Phys. {\bf A20} (1987) 5677.
\end{references}
\end{document}